\documentstyle[12pt]{article}
\begin{document}

\title {MORE ON THE THOMAS PRECESSION AND THE RELATIVISTIC TRANSFORMATION TO ROTATING FRAMES}
\author{L. Herrera\thanks{Postal address: Apartado 80793, Caracas 1080A,
Venezuela; E-mail address: laherrera@movistar.net.ve}
 \\
Escuela de F\'{\i}sica. Facultad de Ciencias.\\ Universidad Central de Venezuela. Caracas, Venezuela.\\
}
\date{}
\maketitle

\begin{abstract}
We comment on some misunderstandings exhibited in a recent paper by Matolcsi et al. \cite{Mato}.
\end{abstract}

\newpage

The Thomas precession and the relativistic definition of a rotating observer are two related issues, which for about a century have been the source of confusions
 and misunderstadings. In this short note we would like to comment on a recent paper illustrating that situation \cite{Mato}.

The Thomas precession  refers to the precession of a gyroscope along an 
arbitrary path in Minkowski space. It was originally calculated by Thomas 
\cite{Th}, when discussing the orbital motion of a spinning electron 
in an atom. We shall consider here the particular case of a circular orbit.

Thus, the Thomas precession is caused by the fact that  the gyroscope is attached to a congruence of rotating observers.
Therefore it is important to keep in mind, that any calculation of the Thomas precession  involves explicitly (or implicitly) the very definition
of the rotating observer. Different definitions {\em must} lead to different results for Thomas precession, independently 
on the approach used to perform such calculations. 

Accordingly, and contrary to what is asserted in \cite{Mato}, there is nothing paradoxical in obtaining different values of precession, when using different congruences of observers
 (each at a time being considered as {\em the congruence associated with rotating observers}). Quite on the contrary, the paradoxical (unacceptable) result would be that different 
congruences of observers, defining rotating observers in (physically) different ways, measure the same precession.

Now, since the early days of special relativity, the well known set of equations
\begin{equation}
t'=t \qquad;\qquad \rho'=\rho \qquad;\qquad 
\phi'=\phi-\omega \,t \qquad;\qquad z'=z 
\label{gal}
\end{equation}
has been adopted, relating the non-rotating coordinate system of the frame $S$ 
(in cylindrical coordinates) to the coordinates $t', \rho', \phi', z'$ of the 
frame $S'$ rotating uniformly about $z$-axis on the $\rho, \phi$ plane.

We may interpret  (\ref{gal}) (hereafter referred to as GAL) as a coordinate transformation.
Obviously, this or any other regular transformation is acceptable, and any physical experiment whose result is expressed through invariants, should yield the same answer
for any observer. However, when we assume that GAL defines the congruence of rotating observers (the congruence of world--line of observers at rest in  $S'$), we are adopting  a precise
 definition of a rotating observer. 
 In other words, one thing is using GAL (which always can be done) and another,
 quite different thing is using GAL and to assume that $S'$ defines a uniformly rotating frame  about $z$-axis on the $\rho, \phi$ plane. Therefore when we say that something 
is rotating, we must specify what we mean by this. This may be done by specifying the congruence of rotating observers 
(e.g. using GAL).

Some problems  with GAL (understood as defining the congruence of rotating observers)
renders it questionable at the ultrarelativistic regime. In order to overcome this situation, a different transformation law 
was independently proposed by Trocheris \cite{Tr} and Takeno \cite{Ta}.

In cylindrical coordinates the Trocheris-Takeno (TT) transformation reads
$$
\rho' = \rho \qquad ; \qquad z' = z
\label{roztt}
$$
\begin{equation}
\phi' = \phi \, \left(\cosh{\lambda}\right) - 
t \, \left(\frac{c}{\rho}\, \sinh{\lambda}\right)
\label{tt}
\end{equation}
$$
t' = t \, \left(\cosh{\lambda}\right) - \phi \, 
\left(\frac{\rho}{c} \, \sinh{\lambda}\right)
$$
with
$$
\lambda \equiv \frac{\rho \omega}{c}
$$
where primes correspond to the rotating frame and $c$ and $\omega$ denote the velocity of light and the  the angular velocity of the rotating frame defined by GAL, respectively .

It can be easily seen \cite{Tr},\cite{Ta}, 
that (\ref{tt}) leads to the relativistic law
of composition of velocities and yields for the velocity of a fixed 
point in $S'$, the expression
\begin{equation}
v=c\tanh{\lambda}
\label{v}
\end{equation}
sending to infinity the ``light cylinder'' (the horizon). A modification of this transformation  (MTT) was proposed in \cite{He}, to ensure that 
rotating observers measure greater circles  as compared with the non-rotating observer.

However we must emphasize that the issue here is not  about the merits or drawbacks of these or any other definition of the congruence of rotating observers. The point is that the
vorticity of different congruences will in general be different, and this will lead to different values of Thomas precession.

Indeed, the vorticity vector, 
is given by
\begin{equation}
\omega^\alpha = \frac{c}{2 \sqrt{-g}} \, \epsilon^{\alpha\beta\gamma\delta} 
u_\beta \omega_{\gamma\delta} = 
\frac{c}{2 \sqrt{-g}} \, \epsilon^{\alpha\beta\gamma\delta} 
u_{\beta} u_{\delta,\gamma}
\label{vv}
\end{equation}
where the vorticity tensor is given by
\begin{equation}
\omega_{\alpha\beta} = u_{\left[\alpha;\beta\right]} - \dot{u}_{[\alpha}u_{\beta]}
\label{vt}
\end{equation}
and $u_{\beta}$ denotes the four vector  tangent to the congruence.

It is important to stress that definitions above applies to {\em any} congruence.

It can be shown \cite{HD} that for MTT and TT congruences, we get
\begin{equation}
\Omega\equiv\left(-\omega_\alpha \omega^\alpha\right)^{1/2} = 
\frac{c}{2\rho} \left(\sinh{\lambda}\cosh{\lambda} + \lambda\right)
\label{Ott}
\end{equation}
whereas for GAL we obtain
\begin{equation}
\Omega = \frac{\omega}{1 - \frac{\omega^2 \rho^2}{c^2}}
\label{Og}
\end{equation}
which is a known result \cite{RiPe}.

Next, since $\Omega$ (for {\em any} time--like congruence) measures the rate of rotation with respect 
to proper time of world lines of points at rest in $S'$, 
relative to the local compass of inertia, then $-\Omega$ describes 
the rotation of the compass of inertia (the ``gyroscope'') with 
respect to reference particles at rest in $S'$ (see \cite{RiPe} 
for detailed discussion on this point). Therefore, after one 
complete revolution the change of orientation of the gyroscope 
as seen by the observer in $S'$, is given by
\begin{equation}
\Delta \phi' = - \Omega \Delta \tau'
\label{or}
\end{equation}
where $\Delta\tau'$ is the proper time interval (in $S'$) 
corresponding to the period of one revolution.

Since  $\Omega_{TT}=\Omega_{MTT}$ and furthermore, the amount of proper time corresponding to a time interval $\Delta t$ in $S$, is the same in both
 $TT$ and  $MTT$ frames, then the change of orientation of the gyroscope for a fixed amount of proper time as measured by each of the (rotating) observers ($TT$ and $MTT$) is the
same, but different from the value obtained for the GAL congruence (as it should be!).

To summarize: 
\begin{itemize}
\item There is not (so far) a unique definition for the congruence of relativistic rotating observers.

\item In order to describe the Thomas precession in terms of physically meaningful quantities, we have to  adopt a specific definition for the congruence of  rotating observers.

\item  Different  definitions  lead to different values of precession. 

\item  Since  there should be only one definition of a  rotating observer, and it should comply with observational evidence, it is clear that 
only the  experiment could tell apart the correct definition of a rotating observer. 

\item The accuracy of such an experiment should be of the order $(\frac{v}{c})^2$ or higher.

\end{itemize}

\end{document}